# A REVIEW OF BLENDED LEARNING IN CONTRAST WITH TRADITIONAL LEARNING IN GHANAIAN UNIVERSITIES

*S. Alhassan*
National Research Tomsk Polytechnic University
*alhassan@tpu.ru*

Blended learning methods constructively fuses more than one delivery mode. It makes use of technology to meet the exigency of the times. Covid-19 pandemic has also strengthened the need for teaching and learning to be in modes other than face-to-face interactions. In this research, a state-of-the-art review is conducted on the impact of blended learning in three universities in Ghana. These universities are selected based on a number of reasons including the use of the uniRank ranking system. It has been shown that the Learning management system is implemented in two out of three universities considered in this research. The study shows that 80% of distance education is conducted in the blended learning mode. It is recommended to target the postgraduate level and subsequently the bachelor's level.

## 1. Introduction

The traditional face-to-face teaching methods have for many years been the main media of transferring knowledge. Curriculum design usually influences the teaching style. The three categories of curriculum design: Forward, central and backward design implement aspects of blended leaning. The educational institutions now have seen transformation over the years which previously focused on bringing secondary education and higher education into closer interaction with the occupational systems of industrial era [1]. Today, due to the incidence of the covid-19 pandemic, the educational institutions have devised innovative ways of making higher education easily accessible. Blended learning has now provided the opportunity to people all around the world, access to quality education regardless of their location. The Covid-19 pandemic, has demonstrated the importance of blended learning in its approach and practice. For Ghana, it under-pinned the cracks in the educational system as it laid bare the very problems the universities faced all this while.

This research is a state-of-the-art review on blended learning and traditional learning in Ghana seeking to influence policy on higher education in the country hence bridging the gap in the implementation of online and face-to-face model into teaching and learning at all universities. In this research, three universities are analyzed to ascertain the impact of Blended learning on teaching and learning.

The International Association of Universities (IAU) is an independent, bilingual, non-governmental organization recognized worldwide as the leading global association of higher educational institution formed under the auspices of UNESCO [2]. The IAU conducted a research survey based on 424 full replies from unique higher education institutions (HEIs) in 109 countries and two special administrative regions of China (Hong Kong and Macao) revealed 77% of universities in Africa, 55% in Asia and the Pacific, and 54% of universities in Europe were closed as a result of the pandemic [3]. Over 90% of these institutions in the various jurisdictions indicated that teaching and learning was adversely affected. Another survey conducted by the organization of economic co-operation and developments' (OECD) on 165 students from 21 African countries indicated that face-2-face interactions were discontinued [4]. 55 out of 69 universities after the covid-19 pandemic have in-cooperated blended leaning in their curricula. This has improved accessibility to the university by students far and near while optimizing the program development cost and time as blended learning has shown a great deal of effectiveness in combining virtual learning sessions or recorded sessions with assignment leading to producing maximum results from students and the faculty [3].

The issues of having the blended learning structure formed, access to internet for such a platform to be uninterrupted and active, and also the economic challenges involved with its implementation remain a challenge to most privately run universities in the country unlike the public ones. Of the about 42 privately run universities in Ghana, less than 5% embraced remote learning leaving the overwhelming majority with the choice of closing down the universities or practicing other protocols of the pandemic [5].

Hence, this research is conducted to analyze the impact of Blended learning in the three universities in order to unearth the challenges, aspects of blended learning implemented so far and the successes attained.

*Blended learning.* is the learning approach where the instructional delivery by the instructor is via technology or even field trips. Content delivery is termed as supports provided both in receiving and applying the data during the teaching and learning process [6]. Based on the method of content delivery and feedback, Blended learning may be categorized into three types, namely: hybrid, flipped classroom or hyflex [7]. In the case of hybrid, online learning activities are designed to replace a portion of the face-to-face learning activities. In the flipped classroom, usually the presentations and lectures are delivered online while the face-to-face interaction is reserved for homework activities. This is usually the case in most humanities and social science programs. For hyflex type, the student is given the liberty to choose which mode to join thus online or face-to-face.

Technology is applied in the form of organizing small online group discussions, online pre-test quiz, reflection essays, virtual classrooms, virtual labs, mini-lectures, flash animation, audios, videos and online. The implementation of blended learning ensures that principles such as the following are attained [8];
a. Improved student to faculty contact
b. Enhanced co-operation amongst students.

c. Timely feedbacks and active learning.

d. Encourage innovation in ways of learning.

The university of Ghana (UG) and the Kwame Nkrumah University of Science and Technology (KNUST) use mostly emails as means of communication. The use of personal webpages, clouds and other announcement outlet is key to improving communication.

*Traditional learning* is defined as the teacher's personal behaviors and media used in transmitting data or receiving it from the learners. The face-to-face approach has been practiced from the inception of teaching and learning. In the classical teaching style, the syllabus with the content is prepared prior with assessments [9]. The instructor is expected to teach exactly the content of the syllabus and conduct assessments provided therein. There is little to no use of technology as all forms of teaching and learning is conducted in a confined location such as a classroom and the students take instructions from the instructor. The only form of assessment is through producing the content thought by the instructor some weeks in the past.

*Higher education* in this research, refers to higher education at the level of universities in Ghana. Institutions of higher education have been established to provide competencies to professionals in diverse careers. The nation has a number of universities with the major direction of studies in Engineering, health sciences, social science and business administration. Depending on the direction of studies the student population could be very large. Only three (3) major universities considered in this research have implemented blended learning [10].

## 2. Universities in Ghana considering Blended learning

Ghana is a West African country located on the coast of the Gulf of Guinea, referred previously to as gold coast due to the enormous amount of that natural resource [11]. With a population of about 32 million, it boosts of about 69 universities excluding other tertiary institutions. The university of Ghana, is largely known as the premier and oldest university in the country, founded in 1948. The Kwame Nkrumah University of Science & Technology (KNUST) established by the first President of the country, Dr. Kwame Nkrumah who is a winner of the Lenin Peace Prize focuses more on Engineering and Health sciences programs. The University of Professional Studies Accra (UPSA) was established to provide professional competencies to actively working stuff of various departments in the government and privately run firms in the country. Earlier, teaching and learning was primarily conducted over the weekends and in the evenings. This is not the case now.

The table 1 below shows the list of 3 universities in the country and their rankings from AD scientific index and QS world rankings [12].

Table 1. Top Ghanaian University Ranking and QS WORLD

| No. | Name of Universities | Ranking in Ghana 2022 | Ranking in Africa 2023 | QS World University Ranking 2023 |
|---|---|---|---|---|
| 1 | University of Ghana (UG) | 2 | 41 | 1001-1200 |
| 2 | Kwame Nkrumah University of Science & Technology | 3 | 41 | 1001-1200 |
| 3 | University of Professional Studies Accra (UPSA) | 22 | 421 | 9786 |

In this paper, the research is conducted to ascertain the impact of blended learning in three universities in Ghana. A number of factors influence the choice of university for the analysis. These include; the student population, the programs or courses available for enrolment and location. Large student population leads to a high non-completion rate. Research on these universities after the advent of covid-19 shows some level of continuity in using the remote learning platform [13].

The rankings as seen in table 1 are based on the uniRank university rankings. The assessment is based on the use of an algorithm with 4 unbiased and independent web metrics extracted from three different web intelligence sources.

## 3. University Education Policy & Blended learning

From the analysis on all three universities in Ghana, the University of professional studies (UPSA) has implemented aspects of blended learning framework in teaching and learning. The university utilizes the learning management system (LMS) [14]. The application of blended learning framework at the UPSA for postgraduate studies in Business Administration achieves 70% online learning activity and interactions. The blended learning program is run, a course model at a time, with a time frame of 5 weeks [14].

The university of Ghana (UG) employs blended learning with use of the learning management system referred to as Sakai. Blended learning is also introduced in programs run under the distance learning model at satellite campus covering 80% of distance education students of over 7000 students [15]. About 5600 students learn using blended learning at the distance education department. These programs make use of the flipped form of the blended learning with content delivery at 90% online while quizzes and final assessments are organized face-2-face [15].

From the analysis, the Kwame Nkrumah University of Science and Technology (KNUST) has demonstrated interest in developing the distance learning programs ranging from the bachelor's level to Postgraduate level without a leaning management system (LMS) [16]. This conclusion is based on the infrastructural development the institution has invested and the capacity of institution. The blended learning format is conducted via virtual classroom accounting for almost 90% of the classroom activity [16]. The replacement form of the blended learning is more suited to the approach implemented there.

Table 2. Indicators of the 3 universities in Ghana considered for Blended learning.

| Key Indicators | Names of Universities | | |
| --- | --- | --- | --- |
| | UG | KNUST | UPSA |
| Geographical Location | Accra | Kumasi | Accra |
| Student population | 53,643 | 85,000 | 14,147 |
| Public/Private University | Public | Public | Public |
| Blended Learning (BL) Metric | | | |
| Learning Management system | Yes | No | Yes |
| Moodle teaching | Yes | No | Yes |
| Instructor personal webpages | No | No | No |
| Video recording of content | Yes | No | No |
| Flash animation | No | No | No |
| Smart/ Virtual classrooms | Yes | Yes | Yes |
| E-learning tools | Yes | Yes | Yes |
| Online assignments/ Grading/ poll | Yes | Yes | Yes |
| Live chats | Yes | No | No |
| Online learning resources | Yes | Yes | Yes |
| Level of online distance learning (%) | 80 | 90 | 70 |
| Virtual labs | No | No | No |
| Video-conferencing rooms | Yes | Yes | Yes |
| Podcast | Yes | No | No |

From table 2 above, the three universities are compared based on the key indicators used in assessing the implementation of blended learning at the various institutions. The blended learning metric shows whether or not the different activities are conducted, and in some instances the extend to which the activity is implemented.

## 4. Conclusion

The research analyzes blended learning methods in contrast to traditional learning establishing the relevance and impact of blended learning in three Ghanaian universities. This research has provided quantitative evidence on the successful implementation of blended learning in higher educational institutions in Ghana post covid-19. It has been shown that the learning management is implemented in two out of three universities considered. The research shows that 80% of distance education in the University of Ghana is conducted in the Blended learning format. At the Kwame Nkrumah University of Science and Technology, 90% of the distance education is by virtual class room while 70% of online learning activities for postgraduate studies in Business Administration is conducted at the University of Professional Studies Accra (UPSA). The evaluation of the key indicators shows that the implementation of LMS, moodle teaching, creating personal webpages of instructors and virtual labs will improve the quality of blended learning in Ghanaian universities. This will lead to achieving full blended at the postgraduate level in the three universities in less than 2 academic years.

## 5. Acknowledgement